\documentclass[12pt]{iopart}
\usepackage{iopams}

\def\ra{{\rm a}}
\def\cH{{\cal H}} \def\cN{{\cal N}} \def\cR{{\cal R}}  
\def\bS{{\bf S}}
\def\bmB{\mathbf{B}}    

\def\CC{\mathbb{C}}\def\RR{\mathbb{R}}
\def\Tr{{\rm Tr}}
\def\d={\buildrel \rm def \over =}
\def\rhoeq{\buildrel \rho \over \sim}

\def\ket#1{\mid~\!\!\!{#1}~\!\!\rangle} \def\bra#1{\langle~\!\!{#1}~\!\!\!\mid}
\def\average#1{\langle~\!\!{#1}~\!\!\rangle}

\begin{document}\jl{1}
\title{Mixed-state twin  observables}
\author{F Herbut\footnote[1]{Permanent address: The Serbian Academy of Sciences and Arts,
Knez Mihajlova 35, 11000 Belgrade, Yugoslavia; E-mail:
fedor@afrodita.rcub.bg.ac.yu} and M Damnjanovi\'{c}}
\address{Faculty of Physics, University of Belgrade, POB 368,
Belgrade 11001, Yugoslavia, http://www.ff.bg.ac.yu/qmf/qsg\_e.htm}

\date{12 January}

\begin{abstract}
Twin observables, i.e. opposite subsystem observables $A_+$ and $A_-$
that are indistinguishable in measurement in a given mixed or pure
state $\rho$, are investigated in detail algebraicly and
geometrically. It is shown that there is a far-reaching
correspondence between the detectable (in $\rho$) spectral entities
of the two operators. Twin observables are state-dependently
quantum-logically equivalent, and direct subsystem measurement of one
of them {\sl ipso facto} gives rise to the indirect (i.e. distant)
measurement of the other. Existence of nontrivial twins requires
singularity of $\rho$. Systems in thermodynamic equilibrium do not
admit subsystem twins. These observables may enable one to simplify
the matrix representing $\rho$.
\end{abstract}

\submitted 

\section{Introduction}
Quantum correlations are one of the most peculiar and amazing
physical idea  underlying quantum theory. They have been attracting
much attention during the development of quantum mechanics. They are
the corner stone for quantum measurement theory, which, applied to
composite systems, makes transparent the conceptual background of
quantum properties, revealing subtle interrelations between the
subsystems \cite{NEUMAN}. The remarkable possibility to predict a
property of one of the subsystems on account of  the result of the
measurement performed on the other one, even when the subsystems are
distant \cite{FV76}, is a substantial ingredient of modern theories
of teleportation \cite{TELE} and quantum computers \cite{QC}.
Nevertheless, most of the existing results apply to pure states of
the composite system, reducing the scope of and probably making
harder the potential applications. The main goal of this paper is to
give several results extending the known properties of the pure state
case \cite{FV76,VF84} to the general (i.e. mixed or pure) state,
which is described by the statistical operator (density matrix).

In the pure state case the subsystem correlations are transparently
sublimated in the notion of the {\em twin observables}. Let the
quantum system $\bS$ be composed of two subsystems $\bS_+$ and
$\bS_-$ (in short: $\bS=\bS_++\bS_-$); the corresponding state space
being  $\cH=\cH_+\otimes\cH_-$, where $\cH_s$ ($s=\pm$) are the state
spaces of the subsystems. When $\bS$ is prepared in a pure state
$\ket{\Phi}\in\cH$, then the partial traces of the projector
$\rho=\ket{\Phi}\bra{\Phi}$ (over the opposite subsystem spaces) give
the statistical operators (mixtures of the second kind \cite{2MIX}):
\begin{equation}\label{Esubstate}
\rho_\pm\d=\Tr_\mp\rho.
\end{equation}
Two opposite-subsystem observables $\{A_+,A_-\}$ are called {\em twin
observables} or {\em twins} if the measurement of $A_s$ in
$\ket{\Phi}$ amounts to the same as the measurement of $A_{-s}$ in
$\ket{\Phi}$. It turned out that for each $s$-subsystem observable
$A_s$ (acting in $\cH_s$) compatible with $\rho_s$, there exists its
twin (opposite-subsystem) observable $A_{-s}$, compatible with
$\rho_{-s}$. More precisely, both twin observables are uniquely
defined only in the {\em relevant subspaces} $\cR_\pm$, i.e. in the
ranges of $\rho_\pm$. The subsystem states $\rho_\pm$ have equivalent
spectral forms unless possibly different defects (in particular,
their ranges are equally dimensional). As a consequence, there exists
(a $\ket{\Phi}$-dependent) anti-unitary mapping $U_\ra$ of $\cR_+$
onto $\cR_-$ such that the restrictions of the twins to the
corresponding relevant spaces $\cR_\pm$ are $U_\ra$-equivalent:
$A_-=U_\ra A_+U^\dagger_\ra$. Since the states in the null spaces of
$\rho_s$ are undetectable, this equivalence manifests itself as
equivalence of the measurements of the twins with respect both to the
obtained results (mean values) and to the post-measurement state
(collapsed by the L\"uders formula, \cite{LUDERS}). In fact, the
twins are indistinguishable by their action on the state
$\ket{\Phi}$, $A_+\ket{\Phi}=A_-\ket{\Phi}$, or equivalently
\begin{equation}\label{Etwindef}
A_+\rho=A_-\rho.
\end{equation}
Hereafter $A_+$ stands also for the operator $A_+\otimes
\mathbf{1}_-$ in $\cH$ ($\mathbf{1}_-$ is the identity in $\cH_-$),
etc.

The scope of definition \eref{Esubstate} is quite general, i.e. it
applies also to the mixed (composite)  states, when the statistical
operator $\rho$ is not a projector. Analogously, the notions of the
subsystem measurements and subsystem observables are by no means
restricted to the pure states. Still, in general, unlike in the pure
state case, the subsystem states $\rho_s$ are not simply related. In
fact, the question of twins, i.e. mutually related opposite-subsystem
observables, seems to be completely open.

In \sref{Stwindef} we begin our investigation by analyzing possible
physically based $\rho$-dependent criteria for equivalence of a pair
of opposite-subsystem quantum events. Subsequently, we generalize
them regaining \eref{Etwindef} as an  unambiguous definition of twin
observables. The importance of the null spaces is emphasized in
\sref{Snull}: when that of $\rho$ is trivial only trivial twins
exist, whereas the null space of the difference of the twins
$(A_+-A_-)$ contains the range of $\rho$. As it is shown in
\sref{Sdetect}, indistinguishability in measurement has physical
contents only for the reducees of the  twins in the ranges of the
subsystem states \eref{Esubstate}. Consequences of the specific
spectral properties of twins are analyzed in the the next two
sections.

We conclude this article by a discussion of the physical meaning of
twin observables within the framework of distant measurement.
Although in the general case the mentioned antiunitary mapping $U_a$
does not exist, the equivalence of the twins in measurement in $\rho$
still holds true. As to {\em practical motivation} for this study, it
is twofold:
\begin{description}
\item{(i)} Perfect correlations between the physically relevant parts of
twins will be established, and they are of interest when the
$\rho$-imposed statistical connection between the two subsystems is
of importance, like  e.g., in quantum information theory.
\item{(ii)} It turns out that the important problem of finding subsystem bases
in which $\rho$ is represented by a matrix that is as simple as possible can be
solved with the help of twins.
\end{description}

\section{Opposite-subsystem equivalent events and twins}\label{Stwindef}
To generalize the notion of the twin observables for the given {\em
general}, i.e. mixed or pure, quantum state (statistical operator or
density matrix) $\rho$, we start with events (projection operators).
At first we note that there are three seemingly different criteria
(two of them physically based) which can be taken to define the
equivalence of two events $E$ and $F$ in the state $\rho$.

Concerning their measurements, these events are {\it observationally
indistinguishable} in $\rho$ if their probabilities $\Tr E\rho$ and
$\Tr F\rho$ are equal and the (unnormalized) states resulting when
the event occurs (gives result $1$) in ideal measurements (i.e.
the collapsed states \cite{LUDERS}) are same:
\begin{equation}\label{1}
E\rho E=F\rho F.
\end{equation}
Actually, this relation implies the equality of the probabilities.

In the special case we are interested in, when $E$ and $F$ are
compatible (commuting) events with non-vanishing probabilities, eq.
\eref{1} {\em implies}
\begin{equation}\label{2}
\Tr(F\frac{E\rho E}{\Tr E\rho})=\Tr(E\frac{F\rho F}{\Tr F\rho})=1.
\end{equation}
This reveals the second criterion: the events $E$ and $F$ are {\em
state-dependently equivalent}, $E\rhoeq F$, implying each other in
$\rho$ in a strengthened quantum-logical sense \cite{F94,F96a,F96b}.

Notice that $E\rhoeq F$ if and only if \cite[Remark 1]{F96a} both the
conditions $E^{\perp}FR=E^{\perp}F$, and $F^{\perp}R=EF^{\perp}$ are
fulfilled ($R$ being null projector of $\rho$). Multiplying them by
$\rho$ from the right, one obtains $F\rho=EF\rho$ and $E\rho
=EF\rho$, yielding the third, algebraic criterion:
\begin{equation}\label{3}
E\rho = F\rho.
\end{equation}

Since \eref{3} obviously implies \eref{1}, all these criteria are
equivalent. They provide us with a satisfactory physical idea of
equivalence or indistinguishability defining twin events. Relation
\eref{3} is very simple and hence it is preferable for further
analysis (and it is a special case of \eref{Etwindef}).

The physical meanings of \eref{1} and \eref{2} are objectionable,
because they are based on ideal measurement (underlying the
L\"{u}ders collapse formula), which is almost quite unperformable in
the laboratory. Hence it is not worth exploring the equivalent
expression \eref{3} in its generality. Still, when a two-subsystem
composite system in the state $\rho$ is considered, with
$E=P_+$ and $F=P_-$ being {\em opposite-subsystem events}, the
mentioned objection may not be applied to \eref{3} having the form:
\begin{equation}\label{4}
P_+\rho=P_-\rho.
\end{equation}
Indeed, because {\em any measurement} (repeatable or unrepeatable) of
$P_s$, not just its ideal measurement, if $P_s$ occurs, leads
\cite[App. 1]{F94} to one and the same {\em opposite-subsystem
conditional state} $\Tr_s\rho P_s/\Tr P_s\rho$. In this case
state-dependent implication does have empirical meaning and \eref{4}
is well worth investigating.

Having thus properly established \eref{4}, we now   regain the
starting relation (2)for twin observables. In fact, the underlying
empirical idea of {\em indistinguishability in measurement} of the
observables $A_\pm$ (in the state $\rho$) amounts now to analogous
indistinguishability of {\em all spectral measures} (projectors)
$P_\pm(\bmB)$ of $A_\pm$ for arbitrary Borel subsets $\bmB$ of
$\mathbb{R}$. Due to the equivalence of \eref{1} and \eref{3}, we
actually have $ P_+(\bmB)\rho=P_-(\bmB)\rho$ for any Borel set
$\bmB$. This, together with the well known functional dependence
$A_\pm=\int_\mathbb{R}\lambda\,\mathrm{d}E_{\lambda} (A_\pm)$
(Stieltjes integral) of the observable on its spectral measure
$E_{\lambda}(A_\pm)\equiv P_\pm(-\infty,\lambda]$, yields the claimed
relation \eref{Etwindef}.

\section{Role of the  null spaces of $\rho$ and $(A_+-A_-)$\label{Snull}}
In the general state case of a composite system the compatibility of
$A_s$ with $\rho_s$ {\em does not} guarantee the existence of its
opposite-subsystem twin $A_{-s}$ as in pure state case. In general,
the problem: "Given the state $\rho$, what are all its twins?" is not
easy to solve, and we do not try to do it in this article.
Nevertheless, the presented  study of twins gives quite general
conditions for their existence.

The basic insight may be obtained when \eref{Etwindef} is rewritten
in the form
\begin{equation}\label{8}
(A_+-A_-)\rho=0.
\end{equation}
It immediately follows that the range of $\rho$ is in the null space
of $(A_+-A_-)$. Hence, if $\rho$ is {\it nonsingular}, $A_+$ and
$A_-$ must coincide, and then the only twins are {\em trivial} (equal
scalar operators): $A_+\otimes\mathbf{1}_-=\mathbf{1}_+\otimes A_-$,
implying $A_s=c\mathbf{1}_s$. More precisely, in a composite-system
state $\rho$ {\em there exist nontrivial twin observables $A_\pm$
only if $\rho$ is singular}. It is noteworthy that this is analogous
to the fact that the state-dependent equivalence $E\rhoeq F$ can be
{\em nontrivial}, i.e. not necessarily just $E=F$, {\it if and only
if} $\rho$ is {\it singular}. Thus, it is the possibly nontrivial
null space of $\rho$ that gives the new quantum logical relations
\cite[Corollary 2]{F94}, as well as possible nontrivial twins.

Note further that \eref{8} is a special ($a=0$) case of the relation of
the type $A\rho=a\rho$, with $A$ Hermitian,  $\rho$ a statistical
operator, and $a$ real. As shown in \ref{A2}, this has a precise
physical meaning: {\it the measurement of the observable $A$ in the
state $\rho$ gives the result $a$ with certainty}.

Let us return to the fact that (7) implies that {\it a sufficient and
necessary condition} for the pair $A_\pm$ to be twins with respect to
$\rho$ is that {\it the range of $\rho$ is  within the null space of
$(A_+-A_-)$}. This has four immediate consequences:
\begin{description}
\item{C1:} All the twins of $\rho$ are also the twins of any
state vector $\ket{\Phi}$ from the range $\cR$ of $\rho$. In fact,
the same is valid for the space $\overline{\cR}$ topologically
closing $\cR$: the null space of $(A_+-A_-)$, as every characteristic
subspace, is topologically closed (because the Hermitian operators
are closed even if they are unbounded), and therefore it contains
$\overline{\cR}$. (This claim is stronger than the preceding one if
the range of $\rho$ is infinitely dimensional and not both twins are
bounded.)
\item{C2:} In any decomposition $\rho=\sum_iw_i\ket{\Phi^{(i)}}\bra{\Phi^{(i)}}$
of $\rho$ into pure states, at least all the twins of $\rho$ itself
are twins of each $\ket{\Phi^{(i)}}$. (This follows from C1.)
\item{C3:}
The set of all $\rho$-twins is the intersection of the sets of all
twins of the pure states contained in $\cR$. Consequently, all
composite states with the same range have the same twins;
equivalently, this means that the $\rho$-twins are completely
specified by the range of $\rho$ only, and are not related to the
finer information contended in the state.
\item{C4:} It may be interesting to get a criterion to recognize the
states $\rho$ admitting given opposite-subsystem observables $A_\pm$
as twins. In fact, $\rho$ is such a state {\it if and only if} it can
be decomposed (like in C2 above) into pure states {\it all being in
the null space of} $(A_+-A_-)$. Moreover, such states can be decomposed
into pure states in no other way.
\end{description}

Finally, \eref{8} also show that examples of twins often may come
from {\it additive observables}. For instance, whenever the
composite-system kinetic energy has the sharp value zero, the
observable $A_+$ can be the kinetic energy of the first subsystem,
while $-A_-$ is that of the second one. In view of this the
conclusion C4 may be helpful in identification of the system state if
such twins have been determined (by experimental evidence). Some
other examples will be considered in \sref{Sctexam}.

\section{The detectable parts of the twins\label{Sdetect}}
We begin now the investigation of {\it twin observables}, i.e., of
two Hermitian subsystem operators $A_+$ and $A_-$ which satisfy
relation \eref{Etwindef}. The subsystem state operators $\rho_\pm$
defined by \eref{Esubstate} will play a role of paramount importance
in our study, since they single out the subspaces essential for the
twin observables concept. This is to a large extent due to their
compatibility with twins (as in the pure state case):
\begin{equation}\label{10}
[A_\pm,\rho_\pm ]=0.
\end{equation}
In fact, using elementary identities for the partial traces and
subsystem operators, and (in the last but one equality) the adjoint
of \eref{Etwindef}, one directly verifies:
\begin{equation*}
A_\pm\Tr_\mp\rho=\Tr_\mp\,A_\pm\rho=\Tr_\mp\,A_\mp\rho=\Tr_\mp\,\rho
A_\mp=\Tr_\mp\,\rho A_\pm=(\Tr_\mp\rho)A_\pm.
\end{equation*}

It is an immediate consequence that the subsystem Hermitian operator
$A_s$ commutes also with all characteristic projectors of the
corresponding subsystem state operator $\rho_s$, and therefore with
any sum of them. In particular, the sum of all characteristic
projectors with positive corresponding characteristic values is the
range projector $R_s$ of $\rho_s$ and the zero characteristic value
projector is null space projector $N_s$. Therefore
\begin{equation}\label{11}
[A_\pm,R_\pm]=0,\qquad [A_\pm,N_\pm]=0.
\end{equation}
These relations imply that the twins reduce in the range $\cR_s$ and
the null space $\cN_s$ of the corresponding subsystem state operator.
We denote by $A'_s$ and $A''_s$ {\it the reducees} in $\cR_s$ and
$\cN_s$ respectively, and we call them the {\it detectable} and the
{\it undetectable parts} of $A_s$, because only the
detectable parts influence measurements (as it will be argued in
\sref{Sspec}). The decompositions
\begin{equation}\label{12}
A_\pm=A'_\pm\oplus A''_\pm\equiv
(A'_\pm\oplus\mathbf{0}''_\pm)~+~(\mathbf{0}'_\pm\oplus A''_\pm )
\end{equation}
thus obtained are, of course, paralleling the decompositions
$\cH_\pm=\cR_\pm\oplus\cN_\pm$ of the subsystem spaces ($\mathbf{0}$
stands for the null operator in the corresponding subspace).

As shown in \ref{A3}, there is a relation among the system and
subsystem states ranges (being accompanied by the equivalent
projector relations):
\begin{equation}\label{13}
\cR\subseteq(\cR_+\otimes\cR_-),\qquad R=RR_+R_-,
\end{equation}
entailing
\numparts\begin{equation}\label{14a}
(\cR_+\otimes\cH_-)\supseteq\cR\subseteq(\cH_+\otimes\cR_-),\qquad
R=RR_\pm=R_\pm R.
\end{equation}
Hence,
\begin{equation*}
\rho=R\rho=R_\pm R\rho=R_\pm\rho,
\end{equation*}
showing that {\it the range projectors} of the subsystem state
operators are {\it always twins}. Note that in the pure state case also
the subsystem state operators $\rho_\pm$ themselves are always twins
 \cite[Eq. (31) and Theorem 8]{FV76}.

Taking the orthocomplements of \eref{14a}, one can write the
following relations for the null spaces $\cN_\pm$, $\cN$ of the
(sub)systems states, and the corresponding projectors $N_\pm$, $N$:
\begin{equation}\label{14b}
(\cN_+\otimes\cH_-)~\subseteq~\cN~\supseteq~(\cH_+\otimes\cN_-),\qquad
N_\pm N=N_\pm.
\end{equation}\endnumparts
Since $N\rho=0$ this yields $N_\pm\rho=0$. Taking into account that
$\mathbf{0}'_\pm\oplus A''_\pm=(\mathbf{0}'_\pm\oplus A''_\pm)N_\pm$,
this entails
\begin{equation}\label{Eundetect}
(\mathbf{0}'_+\oplus A''_+)\rho=(\mathbf{0}'_-\oplus A''_-)\rho=0,
\end{equation}
i.e., the undetectable components are twins (in a trivial way).
Replacing the decompositions \eref{12} in \eref{Etwindef}, and taking
into account the last relation, one obtains
\begin{equation}\label{15}
(A'_+\oplus\mathbf{0}''_+)\rho=(A'_-\oplus\mathbf{0}''_-)\rho,
\end{equation}
i.e. {\it also the detectable components of twins are}, in their
turn, {\it twins}.

Henceforth, the prim on a subsystem entity will denote its
restriction to the subspace $\cR_s$, and the double prim the
restriction on $\cN_s$. On the other hand, the composite system
entities attain prim and double prim when restricted to the subspace
$(\cR_+\otimes\cR_-)$ and its orthocomplement, respectively.

One can strengthen {\it the detectable-undetectable-part aspect of
twins} as follows: Two opposite-subsystem Hermitian operators $A_+$
and $A_-$ are twins {\it if and only if} the following two conditions
are satisfied:
\begin{enumerate}
\item commutation $[A_\pm,R_\pm]=0$, and
\item their detectable parts $A'_\pm$ are twins for $\rho'$.
\end{enumerate}

That these conditions are not only necessary, but also sufficient is,
first of all, obvious from the fact that (i) amounts to the same as
the existence of the two parts of each operator $A_\pm$. Further, the
undetectable parts, whatever they are, are always twins for $\rho''$
because $\rho''=0$ (as the restriction of $\rho$ to a part of its
null space).

>From the analytical point of view, it is not practical to keep the
null spaces of the three state operators in the game (because
everything of interest is trivially zero in them). Henceforth we
mostly restrict the state space $\cH_+\otimes\cH_-$ to its subspace
$\cR_+\otimes\cR_-$. The twin relation \eref{Etwindef} now reduces to
the effective part of \eref{15}:
\begin{equation*}
A'_+\rho'=A'_-\rho'.
\end{equation*}

In accordance with the remark after \eref{8}, if $\rho'$ is {\it
nonsingular}, then {\it there are no nontrivial twins} (in
$\cR_+\otimes\cR_-$). Still, if simultaneously the non-reduced state
operator $\rho$ itself is {\it singular}, then it does have {\it
nontrivial twins}, but these are no other than $\alpha R_\pm$,
$\alpha\in\RR$. (We have utilized the obvious fact that if
$\{A_+,A_-\}$ are twins, so are $\{\alpha A_+,\alpha A_-\}$, $\alpha
\in \RR$).

\section{Characteristic vectors and the characteristic values of the
twins\label{Sspec}}
Considering the commutation relations \eref{10}, one infers that the
twin operators reduce in each characteristic subspace of the
corresponding state operator, and these, except the null space, are
necessarily finite dimensional (because the positive characteristic
values have to add up, repetitions included, into 1). As a
consequence, {\it the spectra of the twin operators} have to be {\it
purely discrete} (in $\cR_+\otimes\cR_-$).

For further study, we confine ourselves  to the subspace
$\cR_+\otimes\cR_-$. Let us take arbitrary {\it characteristic}
orthonormal (ON) bases for the given twins $A'_\pm$ in the given
composite-system state $\rho'$. Denote them by
$\{\ket{m_\pm}~|~\forall m_\pm\}$, and the corresponding
characteristic values by $a^\pm_{m_\pm}$, one obtains the pair of the
characteristic equations
\begin{equation*}
A'_\pm\ket{m_+}\ket{m_-}=a^\pm_{m_\pm}\ket{m_+}\ket{m_-}.
\end{equation*}
When $\rho'$ is applied (from the left) to the both equations, and
the results are subtracted, the adjoint of \eref{8} gives:
\begin{equation*}
0=(a^+_{m_+}-a^-_{m_-})\rho'\ket{m_+}\ket{m_-}.
\end{equation*}
If we assume that $a^+_{m_+}\not=a^-_{m_-}$, then
$\rho'\ket{m_+}\ket{m_-}=0$, i.e.
\begin{equation}\label{16}
\ket{m_+}\ket{m_-}\in
\cN'\qquad\mbox{ if
}a^+_{m_+}\not=a^-_{m_-}.
\end{equation}

Evidently, {\it representing} $\rho'$ in these bases may give
substantial {\it simplification} of the matrix if the twins (or a set
of mutually commuting twins) is chosen with degeneracies of the
characteristic values $a_n$ as small as possible. (See
\sref{Sctexam}) for the best possible case of simplification in this
way.)

To derive a consequence of \eref{16}, let us assume that $a^+$
belongs to the spectrum of $A'_+$, but not to that of $A'_-$. If
$\ket{a^+}$ is a corresponding characteristic vector, then on account
of \eref{16} one obtains for an arbitrary $\ket{\psi}\in\cR_+$:
\begin{equation*}
\bra{\psi}\rho'_+\ket{a^+}=
\sum_{m_-}\bra{\psi}\bra{m_-}\rho'\ket{a^+}\ket{m_-}=0.
\end{equation*}
Since $\ket{\psi}$ is arbitrary, we end up with $\rho'_+\ket{a^+}=0$,
which contradicts our starting assumption. The symmetric argument
goes through analogously.

Thus we conclude that the twins $A'_\pm$ must have {\it equal
spectra}:
\begin{equation}\label{Espec}
\sigma'=\sigma'(A_\pm)=\sigma(A'_\pm).
\end{equation}
Notice that in general the corresponding {\em multiplicities are not
equal}, as clearly illustrated by the example of $cR_\pm$ above,
despite the particular {\em equality in the pure state case}
\cite{FV76,VF84}.

In order to simplify the forthcoming study of the
characteristic projectors, we prove in \ref{Atfun} the following basic
structural properties of the set of all the twins for given composite
state:
\begin{enumerate}\item For an arbitrary (Hermitian)
operator function $F$ on the twins $A_\pm$, the operators $F(A_+)$
and $F(A_-)$ are also twins, i.e.
\numparts\begin{equation}\label{18} F(A_+)\rho=F(A_-)\rho.
\end{equation}
\item The set of pairs of $\rho$-twin observables is symmetric polynomial algebra, i.e.
any (real) symmetric polynomial $F(x,y,\dots)$ maps pairs
$A_\pm,B_\pm,\dots$ of twin observables into twins:
\begin{equation}\label{Epoly}
F(A_+,B_+,\dots)\rho=F(A_-,B_-,\dots)\rho.
\end{equation}\endnumparts
\end{enumerate}
We emphasize the consequence that (Hermitian) twin observables form
real vector spaces, enabling us to define the set by its basis. These
results are also helpful when new twins are to be generated from
known ones.

Now we turn to characteristic projectors $P_\pm(a^\pm)$ corresponding
to the characteristic values $a^\pm$ of the twins $A_\pm$. Note that
\eref{11} implies $[P_\pm(a^\pm),R_\pm]=0$ for any $a^\pm$. Then the
equality
\begin{equation*}
\Tr\,P_\pm(a^\pm)\rho=\Tr\,P'_\pm(a^\pm)\rho'
+\Tr\,P''_\pm(a^\pm)\rho''=\Tr\,P'_\pm(a^\pm)\rho',
\end{equation*}
reveals that the positive-probability characteristic values of twins
$A_\pm$ are those and only those remaining in the spectra of
$A'_\pm$. This means physically that only the spectral events of
$A'_\pm$ are detectable in $\rho$. This  justifies the term
"detectable part".

Moreover, due to the equal spectra of the detectable parts (cf.
\eref{Espec}), the well known spectral projector polynomials are the
same function for both twins:
\begin{equation*}
P'_\pm(a)=\prod_{(a\ne)b\in\sigma'}\frac{A'_\pm-b}{a-b}\qquad\forall
a\in\sigma'.
\end{equation*}
In view of \eref{18}, $P'_+(a)\rho'=P'_-(a)\rho'$ follows. To
conclude, {\em all positive-probability characteristic projectors}
$P'_\pm(a)$, of twins are twins. This is not only necessary, but {\it
also sufficient}, because
\begin{equation*}
A'_\pm\rho'=\sum_{a\in\sigma'}aP'_\pm(a)\rho'.
\end{equation*}
On account of (13), also $P_{\pm}(a)$ , $a\in \sigma'$ are twins, and
$$ A_{\pm}\rho = \sum_{a\in \sigma'}aP_{\pm}(a)\rho .$$

As to the undetectable parts $\sigma''(A_{\pm})$ of the spectra, each
possible pair of the characteristic projectors are also twins, but
annihilating $\rho$, in the trivial sense of \eref{Eundetect}.

\section{Complete twins and examples}\label{Sctexam}
Let $\rho'_{12}$ be such that a pair of twins $A'_\pm$ with {\it all
characteristic values nondegenerate} exists. We call such operators
{\it complete twins}. Since now the spectra of the two operators
completely coincide, i.e., also the multiplicities are equal, their
ranges $\cR_\pm$ are equally dimensional. This is a necessary
condition for the existence of complete twins.

If $\rho'$ does have a pair of complete twins $A'_\pm$ with the
common spectrum $\sigma'$ and corresponding characteristic ON bases
$\{\ket{a}_\pm~|~a\in\sigma'\}$, then due to \eref{16} the matrix
representing $\rho'$ has the form
\begin{equation}\label{19}
\bra{a}\bra{c}\rho'\ket{b}\ket{d}=
\delta_{ac}\delta_{bd}\bra{a}\bra{a}\rho'\ket{b}\ket{b}.
\end{equation}
This is the maximal simplification that one can achieve by using a
pair of twins or a pair of sets of mutually compatible twins.

As mentioned in the Introduction, when the composite system state is
pure, $\rho'\equiv\ket{\Phi}\bra{\Phi}$, each subsystem observable
compatible with the corresponding subsystem state has its
opposite-subsystem twin. Therefore, complete twins now exist
\cite{FV76,VF84}. The vectors in a common characteristic bases of
$\rho'_\pm$ and such an operator  $A'_\pm$,
$\{\ket{a}_\pm~|~a\in\sigma'\}$ (the corresponding characteristic
values $r_a$ of $\rho_\pm$ are the same) are determined up to phase
factors. If these are simultaneously chosen in both subsystem spaces
according to
\numparts\begin{equation}\label{20b}
\ket{a}_-=U_\ra\ket{a}_+\d={\rho'}^{-1/2}_-\bra{a}_+\ket{\Phi},\qquad\forall
a\in\sigma',
\end{equation}
(a partial scalar product on the right), then such bases give the
Schmidt canonical forms (or biorthogonal expansions) of $\ket{\Phi}$:
\begin{equation}\label{20a}
\ket{\Phi}=\sum_ar^{1/2}_a\ket{a}\ket{a}.
\end{equation}\endnumparts

Let
\begin{equation*}
\rho'=\sum_iw_i\ket{\Phi^{(i)}}\bra{\Phi^{(i)}}
\end{equation*}
be a decomposition of the given composite-system state $\rho'$ into
pure states (orthogonal or not), and let the state $\rho'$ have a
pair of complete twins $A'_\pm$. Then,  according to the conclusion
C2 of \sref{Snull}, this is a common pair of twins for all admixed
states $\ket{\Phi^{(i)}}\bra{\Phi^{(i)}}$. It will not necessarily
lead to simultaneous Schmidt canonical forms \eref{20a} because the
conditions \eref{20b} need not be simultaneously satisfiable. But, if
one relaxes the requirement that the biorthogonal expansion have
positive expansion coefficients, then, with an arbitrary pair of
characteristic bases of the twins, one obtains {\it simultaneous
generalized Schmidt biorthogonal expansions}:
\begin{equation}\label{21}
\ket{\Phi^{(i)}}=\sum_a\alpha_a^{(i)}\ket{a}\ket{a},\quad\forall
i\qquad (\alpha_a^{(i)}\in \CC).
\end{equation}
The bases $\ket{a}_\pm$ are characteristic ones not only for $\rho_\pm$,
but also for all pure state subsystem operators
$\rho^{(i)}_\pm=\Tr_\mp\ket{\Phi^{(i)}}\bra{\Phi^{(i)}}$
(the corresponding characteristic values are
$r^{(i)}_a=|\alpha^{(i)}_a|^2$). Consequently, having simultaneously
these characteristic bases, all operators
\begin{equation}\label{22}
A_s,\{\rho^{(i)}_s:\forall i\},\rho_s\qquad (s=\pm)
\end{equation}
are {\em compatible} (separately for each $s$ of course).

A mixed state $\rho'$ with a pair of complete twins can be obtained
by mixing pure states that do have common generalized Schmidt
biorthogonal expansions \eref{21}. Then, any other decomposition of
the composite-system state (the spectral form included), also has
this property.

To conclude the section, we give several examples of complete
twins in the two particle spin spaces. The observable of the single-
particle $z$-component of spin is denoted by $s_{z\pm}$, yielding
quantum numbers $m_\pm$, while $S$ and $M_S$ are the total spin and its
$z$-projection quantum numbers.

{\it Example 1.} The state space of the system of two spin $1/2$
particles is $\CC^2\otimes\CC^2$. When the range of a mixed state
$\rho$ is spanned by the two state vectors $\ket{S=1,M_S=0}$ and
$\ket{S=0,M_S=0}$, all twins have the form
$A_\pm=\alpha\mathbf{1}_\pm\pm\beta s_{z\pm}$
($\alpha,\beta\in\mathbb{R}$). Thus the additive-type complete twins
$\pm s_{z\pm}$, together with (trivial) identities, span the
two-dimensional twin space. Let us note that the (at first sight)
similar example of the states with the range spanned by
$\ket{S=1,M_S=0}$ and $\ket{S=1,M_S=-1}$, admitting only
$\alpha\mathbf{1}_\pm$ twins, illustrates that singularity of $\rho$
is not sufficient for the occurrence of nontrivial twins.

{\it Example 2.} For two spin 1 particles the state space is
$\CC^3\otimes\CC^3$. Whenever $\cR$ is the entire $M_S=0$ subspace
spanned by the three vectors $\ket{S=2,M_S=0}$, $\ket{S=1,M_S=0}$ and
$\ket{S=0,M_S=0}$, the space of twins is spanned by $\mathbf{1}_\pm$,
$s^2_{z\pm}$ and $\pm s_{z\pm}$. Only the last are complete twins
although also the second ones are nontrivial. A more subtle analysis
may be performed for the states with the range $\cR$ equal to the
$M_S=1$ subspace, spanned by $\ket{S=2,M_S=1}$ and $\ket{S=1,M_S=1}$.
Here the single-particle state operators $\rho_\pm$ are singular
(with the null spaces spanned by $m_\pm=-1$ states), in contrast to
the former examples. In fact, the twin space is spanned by
$\mathbf{1}_\pm$, $A_\pm=\pm s_{z\pm}\mp\frac12\mathbf{1}_\pm$ and
$s^2_{z\pm}-s_{z\pm}$. In addition to the first pair, the last one is
also trivial, but in the sense of \eref{Eundetect} as a pair of
undetectable observables. On the other hand, $A_\pm$ are complete
twins, based on the sharply valued additive observable $S_z$ (on
$\cR$) (cf. the final remark in the next section). Their detectable
parts  $A'_\pm=\pm\frac12{1\ \ 0\choose 0\ -1}$ are complete twins;
supplemented by $\mathbf{0}''_\pm$ in the undetectable null spaces
these yield the twins $A'_\pm\oplus\mathbf{0}''_\pm=\mp\frac12
\mathbf{1}_\pm\pm\frac34s^2_{z\pm}\pm\frac14s_{z\pm}$ which together
with the subsystem range projectors
$R_\pm=\mathbf{1}_\pm-\frac12s^2_{z\pm}+\frac12s_{z\pm}$, span the
space of the detectable twins.

\section{Concluding remarks}
The fact that the detectable parts $A'_\pm$ of twin observables in a
composite-system state $\rho$ have common discrete characteristic
values $a\in\sigma'$ with possibly differing multiplicities and no
continuous spectrum has various consequences.

First of all, the physical meaning of twin observables becomes
transparent in terms of distant measurement, when the experimental
indistinguishability of $A_+$ and $A_-$ in $\rho$ is expressed in a
more detailed way as follows. Any result $a$ is obtained with the
same probability irrespectively if $A_+$ or $A_-$ is measured in
$\rho$, and, if the measurement is ideal, it is accompanied by the
same change of state
\begin{equation}\label{23}
\rho\mapsto\frac{P_+(a)\rho P_+(a)}{\Tr\,P_+(a)\rho}=
\frac{P_-(a)\rho P_-(a)}{\Tr\,P_-(a)\rho}.
\end{equation}
Naturally, also the expectation values $\average{A_+}$ and
$\average{A_-}$ are the same in $\rho$.

This fact is, perhaps, more intriguing when put in the following
way. One can {\it measure} $A_-$ indirectly, or, as one says, {\it
distantly}, in $\rho$ as a sheer consequence of the actual direct
measurement of the nearby opposite-subsystem observable $A_+$ in this
state (or {\it vice versa}). When two particles in a correlated state
are literally distant from each other, then this measurement of the
distant second-particle observable is performed in a ghostly way, by
not "touching" dynamically the distant particle, and performing
measurement only on the nearby (first) particle.

The $\rho$-dependent (strengthened) quantum-logical
equivalence of $P_+(a)$ and $P_-(a)$ for a mentioned result $a$ can
be spelled out as follows. If $P_+(a)$ occurs in an arbitrary
measurement, the opposite-subsystem event $P_-(a)$ becomes {\it
certain}, and vice versa. Since this claim does not involve the
global behaviour of the composite system, we can refer to this
property of $P_\pm(a)$ as to {\it local physical twins}.
Contrariwise, having in mind the common {\it global change of state}
\eref{23}, we can can speak of these events as of {\it global
physical twins}.

It turns out that the correlations between the subsystems are closely
connected with the null space of the composite state. Indeed,
physically nontrivial twins exist only for  singular $\rho$:
otherwise, the only twins are $\alpha\mathbf{1}_\pm$
($\alpha\in\RR$), when their measurements are without content.
Consequently, a complex system in equilibrium that is described by
the canonical ensemble state $\rho=\e^{-H/kT}/\Tr\e^{-H/kT}$ cannot
be decomposed into twins-rigged subsystems for $T>0$ (when $\rho$ is
nonsingular). In fact, the conclusion C3 in \sref{Snull} shows that
only geometrical relations among the (sub)system state ranges
$\cR_\pm$ and $\cR$ are relevant for the twins-type correlations.

Nevertheless, the problem of determining the set of all twins for a
given $\rho$ remains hard enough. (The direct solution of the linear
system \eref{Etwindef} is possible only for especially simple, thus
mostly unrealistic, systems). The structural properties stated by
(17) should be helpful.

On the other hand, one should note that the additive observables are
promising candidates for twin observables. In fact, whenever a
composite system is prepared in a  state $\rho$ with the sharp value,
say $b$, of an additive observable $B=B_++B_-$, then the subsystem
observables $\pm B_\pm\mp \frac{b}{2}\mathbf{1}_\pm$ are twins as
follows from \eref{8}. Note that the range $\cR$ is in this case
confined to the corresponding characteristic subspace of $B$, i.e.
$RP(b)=R$, in accordance with the required singularity of $\rho$.

\ack We are thankful for the helpful discussions we have had
in an early stage of work on this problem with our colleagues
professors M. Buri\'{c}, I. Ivanovi\'{c} and M.
Vuji\v{c}i\'{c}.

\appendix

\section{Certainty of $a$ in a state satisfying $A\rho=a\rho$\label{A2}}

Applying the relation $A\rho=a\rho$ to an arbitrary state vector
$\ket{\psi}$ one obtains $A\rho\ket{\psi}=a\rho\ket{\psi}$. This
means that $\rho\ket{\psi}$ is a characteristic vector of $A$ for
the characteristic value $a$, independently of $\ket{\psi}$, i.e.
that the range of $\rho$ is within range of the corresponding
characteristic projector $P_a$ of $A$. Therefore, $P_a\rho=\rho$,
implying that the probability $\Tr P_a\rho$ of the result $a$
measuring $A$ in $\rho$ is $1$.

Moreover, the condition we have started with is also necessary for
the result $a$ in measurement of $A$ in $\rho$ to be certain, because
the entire argument can be read backwards. The only nontrivial step,
i.e. that $\Tr P_a\rho =1$ implies $P_a\rho =\rho$, is proven in
\cite[Lemma A.2]{F94}. Thus, this probability-one statement is the
precise physical meaning of the condition at issue.

\section{Relation among the ranges and null spaces\label{A3}}
For arbitrary normalized $\ket{\psi_-}\in\cH_-$ there is an
orthonormal basis $\{\ket{i_-}~|~\forall i_-\}$ of $\cH_-$ containing
$\ket{\psi_-}$. Then, since $\rho$ is positive, the definition
\eref{Esubstate} gives for each $\ket{n_+}$ from the null space
$\cN_+$ of $\rho_+$:
\begin{equation*}\label{Eker}\fl
0\le\bra{n_+}\bra{\psi_-}\rho\ket{n_+}\ket{\psi_-}\le
\sum_{i_-}\bra{n_+}\bra{i_-}\rho\ket{n_+}\ket{i_-}=
\bra{n_+}\rho_+\ket{n_+}=0.
\end{equation*}
Thus, $\ket{n_+}\ket{\psi_-}$ is in the null space $\cN$ of $\rho$,
i.e. $(\cN_+\otimes\cH_-)\subseteq\cN$; symmetrically:
$(\cH_+\otimes\cN_-)\subseteq\cN$. Besides,
$\cH=(\cN_+\otimes\cN_-)~\oplus~(\cN_+\otimes\cR_-)~\oplus~
(\cR_+\otimes\cN_-)~\oplus~(\cR_+\otimes\cR_-)$, since the ranges
$\cR_\pm$ and $\cR$ of the (sub)system states are orthocomplements of
$\cN_\pm$ and $\cN$ respectively. Finally,  it follows:
\begin{equation}\label{Enull}
\cN\supseteq(\cN_+\otimes\cN_-)~\oplus~(\cN_+\otimes\cR_-)~\oplus~
(\cR_+\otimes\cN_-),
\end{equation}
directly implying \eref{13}.

\section{Proof of structural properties (17) of the set of twins\label{Atfun}}
To prove \eref{18}, let $\{\ket{m_\pm}~|~\forall m_\pm\}$ be
characteristic complete ON bases of the respective range projectors
$R_\pm$ (of $\rho_\pm$), such that their subbases are also
characteristic bases of the respective twins $A'_\pm$. Then we verify
the twin relation in the form of the adjoint of \eref{8} for
$F(A_\pm)$. Indeed,
\[\rho(F(A_+)-F(A_-))\ket{m_+}\ket{m_-}=
(F(a^+_{m_+})-F(a^-_{m_-}))\rho\ket{m_+}\ket{m_-}=0,\]
since $\ket{m_+}\ket{m_-}\in\cN$ whenever $\ket{m_+}\in\cN_+$ and/or
$\ket{m_-}\in\cN_-$ (by \eref{14b}), or $a^+_{m_+}\not=a^-_{m_-}$ (by
\eref{16}), and in the remaining case (i.e. when
$\ket{m_\pm}\in\cR_\pm$ and $a^+_{m_+}=a^-_{m_-}$) obviously
$F(a^+_{m_+})-F(a^-_{m_-})=0$.

As for \eref{Epoly}, a straightforward consequence of \eref{Etwindef}
is that $\alpha A_\pm+\beta B_\pm$ are twins, while the assertion on
the symmetric products follows from:
\[\fl(A_+B_++B_+A_+)\rho=(A_+B_-+B_+A_-)\rho=(B_-A_++A_-B_+)\rho=
(B_-A_-+A_-B_-)\rho\].


\end{document}